%% file: main.tex
\documentclass[twocolumn,showpacs,preprintnumbers,nofootinbib,prd,
superscriptaddress,10pt]{revtex4-1}

\usepackage{lmodern}
\usepackage{amsmath,amssymb}
\usepackage[normalem]{ulem}
\usepackage{textcomp}
\usepackage{hyperref}
\usepackage{bm}
\usepackage{graphicx}
\usepackage{psfrag}
\usepackage{aas_macros}
\usepackage[usenames,dvipsnames]{xcolor}
\usepackage{orcidlink}
\usepackage[utf8]{inputenc}
\usepackage{lipsum} 
\usepackage{multirow}
\usepackage{soul}
\usepackage{booktabs} 

\setcitestyle{etalmode=truncate, maxnames=3, minnames=1}

\definecolor{darkgreen}{rgb}{0,0.5,0}

\hypersetup{
    bookmarks=true,         
    unicode=false,          
    pdftoolbar=true,        
    pdfmenubar=true,        
    pdffitwindow=false,     
    pdfstartview={FitH},    
	pdftitle={Biased parameter inference of eccentric, spin-precessing binary black holes},
    pdfauthor={Divyajyoti et al.},
    pdfsubject={Gravitational Waves},   
    pdfkeywords={gravitational waves, eccentric compact binaries, general relativity},
    pdfnewwindow=true,      
    colorlinks=true,       
    linkcolor=red,          
    citecolor=Blue,        
    filecolor=magenta,      
    urlcolor=Blue,           
    linktocpage=true
}

\allowdisplaybreaks[4]

\DeclareFontFamily{OT1}{pzc}{}
\DeclareFontShape{OT1}{pzc}{m}{it}{<-> s * [1.10] pzcmi7t}{}
\DeclareMathAlphabet{\mathpzc}{OT1}{pzc}{m}{it}

\newcommand{\ICTS}{International Centre for Theoretical Sciences, Tata Institute of Fundamental Research, Bangalore 560089, India}

\newcommand{\IITM}{Department of Physics, Indian Institute of Technology Madras, Chennai 600036, India}

\newcommand{\CSGC}{Centre for Strings, Gravitation and Cosmology, Department of Physics, Indian Institute of Technology Madras, Chennai 600036, India}

\newcommand{\AEI}{Max Planck Institute for Gravitational Physics (Albert Einstein Institute), Am M{\"u}hlenberg 1, 14476 Potsdam, Germany}

\newcommand{\Cornell}{Cornell Center for Astrophysics and Planetary Science, Cornell University, Ithaca, New York 14853, USA}

\newcommand{\Caltech}{Theoretical Astrophysics 350-17, California Institute of Technology, Pasadena, CA 91125, USA}

\newcommand{\Cardiff}{Gravity Exploration Institute, School of Physics and Astronomy, Cardiff University, Cardiff, CF24 3AA, United Kingdom}

\newcommand{\PSU}{Institute for Gravitation \& the Cosmos, Department of Physics, Penn State University, University Park, PA 16802, USA}

\begin{document}

\title{Biased parameter inference of eccentric, spin-precessing binary black holes}


\author{Divyajyoti~\orcidlink{0000-0002-2787-1012}}
\email{divyajyoti.physics@gmail.com}
\affiliation{\Cardiff}
\affiliation{\IITM}
\affiliation{\CSGC}

\author{Isobel M. Romero-Shaw~\orcidlink{0000-0002-4181-8090}}
\affiliation{\Cardiff}

\author{Vaishak Prasad~\orcidlink{0000-0001-6712-2457}}
\affiliation{\PSU}
\affiliation{\ICTS}

\author{Kaushik Paul~\orcidlink{0000-0002-8406-6503}}
\affiliation{\ICTS}
\affiliation{\IITM}
\affiliation{\CSGC}

\author{Chandra Kant Mishra~\orcidlink{0000-0002-8115-8728}}
\affiliation{\IITM}
\affiliation{\CSGC}

\author{Prayush Kumar~\orcidlink{0000-0001-5523-4603}}
\affiliation{\ICTS}

\author{Akash Maurya~\orcidlink{0009-0006-9399-9168}}
\affiliation{\ICTS}

\author{Michael Boyle~\orcidlink{0000-0002-5075-5116}}
\affiliation{\Cornell}

\author{Lawrence~E.~Kidder~\orcidlink{0000-0001-5392-7342}}
\affiliation{\Cornell}

\author{Harald~P.~Pfeiffer~\orcidlink{0000-0001-9288-519X}}
\affiliation{\AEI}

\author{Mark~A.~Scheel~\orcidlink{0000-0001-6656-9134}
}\affiliation{\Caltech}

\begin{abstract}

While the majority of gravitational wave (GW) events observed by the LIGO and Virgo detectors are consistent with mergers of binary black holes (BBHs) on quasi-circular orbits, some events are also consistent with non-zero orbital eccentricity, indicating that the binaries could have formed via dynamical interactions. 
Moreover, there may be GW events which show support for spin-precession, eccentricity, or both.  
In this work, we study the interplay of spins and eccentricity on the parameter estimation of GW signals from BBH mergers. 
We inject eccentric signals with no spins, aligned spins, and precessing spins using hybrids, \textsc{TEOBResumS-DALI}, and new Numerical Relativity (NR) simulations, respectively, and study the biases in the posteriors of source parameters when these signals are recovered with a quasi-circular precessing-spin waveform model, as opposed to an aligned-spin eccentric waveform model. 
We find significant biases in the source parameters, such as chirp mass and spin-precession ($\chi_p$), when signals from highly-eccentric BBHs are recovered with a quasi-circular waveform model. 
Moreover, we find that for signals with both eccentricity and spin-precession effects, Bayes factor calculations confirm that an eccentric, aligned-spin model is preferred over a quasi-circular precessing-spin model. 
Our study highlights the complex nature of GW signals from eccentric, precessing-spin binaries and the need for readily usable inspiral-merger-ringdown eccentric, spin-precessing waveform models for unbiased parameter estimation.

\end{abstract}


\maketitle
\section{Introduction}
\label{sec:intro}

The most recent catalog of gravitational wave (GW) transients released by the LIGO-Virgo-KAGRA (LVK)~\citep{LIGOScientific:2014pky, KAGRA:2020tym, VIRGO:2014yos} collaboration contains 158 confident compact binary merger signals, of which 153 likely originate from binary black hole (BBH) mergers \citep{LIGOScientific:2025slb, LIGOScientific:2025pvj}.\footnote{The 158 mergers quoted here are the subset of the 218 mergers reported in \citep{LIGOScientific:2025slb} that are used to infer population properties in \citep{LIGOScientific:2025pvj}. The latter paper has a stricter threshold for inclusion than the main catalogue (false alarm rate $FAR<1~\mathrm{yr}^{-1}$ as opposed to astrophysical probability $p_\mathrm{astro} \geq 0.5$).} There is increasing evidence from the mass and spin distributions of these binaries that \textit{hierarchical} mergers---systems that contain one or more remnants of previous mergers, which can only occur in densely-populated environments---are present in the data \citep[e.g.][]{Kimball:2020qyd, Antonini:2024het, Li:2025fnf, MaganaHernandez:2025fkm, Antonini:2025ilj, Tong:2025wpz}. Individual events can be pointed to as most likely having formed hierarchically: for example, GW231123, with component masses of $\approx 137$ and $\approx 104$~M$_\odot$ and high component spins, exhibits component characteristics typically expected for merger remnants \citep{gw231123}. Binaries that merge in densely-populated environments form \textit{dynamically}. Hierarchical mergers are a small percentage of the total population that merges in dynamical environments \citep[e.g.,][]{Mapelli:2021syv, Gerosa:2021mno, Torniamenti:2024uxl, Banagiri:2025dmy}; a small sub-population of hierarchical mergers indicates that a larger fraction of the population forms dynamically~\citep{Zevin:2021rtf}.

In addition to hierarchical mergers that have a distinct mass and spin distribution, dynamical environments are predicted to produce binaries with non-negligible orbital eccentricity in ground-based gravitational-wave detectors \citep[e.g.,][]{Rodriguez:2017pec, Zevin:2021rtf, DallAmico:2023neb, Samsing:2020tda}. This is in contrast with \textit{isolated} evolution, through which binaries circularise before they enter the frequency band of the current generation of Advanced-era ground-based GW detectors \citep{Peters:1964zz}. The exception is binaries that evolve as a constituent of isolated triples, which can merge with detectable eccentricities due to the driving influence of the tertiary \citep[e.g.,][]{Rodriguez:2018jqu, Liu:2019gdc, Dorozsmai:2025jlu}. 

In dynamical environments like globular clusters (GCs) and active galactic nuclei (AGN), as well as in field triples, the subset of mergers that have detectable eccentricity close to merger is a small fraction of the total population ($\sim5\%$ to $\sim10\%$) \citep[e.g.,][]{Rodriguez:2018jqu, Rodriguez:2017pec, Zevin:2021rtf, Tagawa:2020jnc}. A confident detection of an eccentric binary would be a smoking gun for non-isolated evolution. There are several claims that some BBH signals detected by the LVK are more consistent with eccentric than quasi-circular orbits: for example, GW190521, GW200129, and GW200208\_22 have all been identified by more than one independent study as possible eccentric mergers \citep[e.g.,][]{Romero-Shaw:2020thy, Gayathri:2020coq, Gamba:2021gap, Romero-Shaw:2022xko, Gupte:2024jfe, Planas:2025jny, Romero-Shaw:2025vbc, McMillin:2025hof}. However, none of these claims are iron-clad: all of these analyses use waveform models that are restricted to aligned BH spins, and the effects of eccentricity have known degeneracies with the effects of misaligned-spin-induced orbital-plane precession \citep{Romero-Shaw:2022fbf, Xu:2022zza, Divyajyoti:2023rht}. Meanwhile, all catalogs of the LVK so far have used waveform models that assume quasi-circularity \citep{LIGOScientific:2018mvr, LIGOScientific:2020ibl, LIGOScientific:2021usb, KAGRA:2021vkt, LIGOScientific:2025slb}.

Improvements in detector sensitivity have facilitated higher detection rates and higher signal-to-noise ratios (SNRs), enabling more detailed and accurate characterisation of both individual binaries and the overall population. However, with increasing SNR, waveform models used for inference are required to include more physics and have greater faithfulness to numerical relativity simulations: in new regions of parameter space, we are now seeing significant differences in parameters recovered with different waveform models \citep[e.g.,][]{gw231123}. Neglecting the effects of orbital eccentricity or spin-precession has already been shown to lead to biases in recovered parameters with detectors at third-observing-run sensitivity \citep{Romero-Shaw:2020thy, Divyajyoti:2023rht}, and with increasing sensitivity we can expect that the impacts and extents of these biases worsen.

Several complete (inspiral-merger-ringdown), cutting-edge eccentric waveform models (including \textsc{TEOBResumS-DALI}~\cite{Nagar:2018zoe} and \textsc{IMRESIGMA}~\citep{Paul:2024ujx} which we have used in this study) containing both eccentricity parameters (eccentricity and mean/relativistic anomaly at the reference frequency) as well as higher-order modes are now available, and efficient enough for use with highly-parallelised or machine learning-assisted Bayesian inference codes \citep[e.g.,][]{Nagar:2024dzj, Gamboa:2024hli, Planas:2025feq}; see also \cite{Klein:2018ybm, Klein:2021jtd, Klein:2013qda, Huerta:2014eca, Moore:2018kvz, Tanay:2019knc, Liu:2019jpg, Tiwari:2020hsu, Albanesi:2023bgi, Albertini:2023aol, Nagar:2024dzj, Nagar:2024oyk, Huerta:2016rwp, Hinderer:2017jcs, Hinder:2017sxy, Cao:2017ndf, Chiaramello:2020ehz, Nagar:2021gss, Ramos-Buades:2021adz, Manna:2024ycx, Carullo:2023kvj, Carullo:2024smg, Yun:2021jnh, Becker:2024xdi, Estelles:2020osj, Cotesta:2018fcv, Ramos-Buades:2019uvh, Islam:2024bza, Islam:2024zqo, Islam:2024rhm, Islam:2025rjl, Islam:2025llx} for other eccentric waveform models, including \textsc{TaylorF2Ecc}~\cite{Moore:2016qxz, Kim:2019abc} used in this study. Recently, new waveform models have been developed that also include the effects of spin-induced precession, although these are limited by being inspiral-only models \citep{2025PhRvD.111h4052M} or by being computationally expensive \citep{2024CQGra..41s5019L, 2025arXiv250314580A}. Simulations of BBH mergers occurring in field triples, GCs, and AGN show that detectably-eccentric mergers from these channels are likely to also have misaligned spins \citep[e.g.,][]{Rodriguez:2017pec, Antonini:2024het, Stegmann:2025shr}. 
It is therefore a matter of urgency that we understand, and are able to mitigate the biases that originate from analysing eccentric \textit{and/ or} spin-precessing signals with waveforms that neglect one of the two effects.

Preliminary attempts have been made to quantify and mitigate the extent of biases that arise from analysing eccentric and/or spin-precessing signals with models that neglect one or both of these effects. \citet{Romero-Shaw:2022fbf} demonstrate that an eccentric or precessing signal can be confidently and correctly identified via comparison of Bayes factors as long as the signal is of adequate length. Xu et al.~\citep{Xu:2022zza} show that for high-mass, short signals, eccentricity lower than 0.2 at $10$~Hz is insufficient to mimic precession. 
\citet{Divyajyoti:2023rht} find that for low-mass systems, low eccentricities do not mimic spin-precession.

In this work, we aim to quantitatively and comprehensively address the effects of eccentricity and spin-precession. We perform injections of eccentric signals in zero-noise, and carry out parameter estimation with varying degrees of spin complexity, using waveforms from different families, to study the effect of spins and eccentricity on the source parameters.

We observe that for all spin configurations, the bias in chirp mass and spin-precession parameter ($\chi_p$) generally increases with an increase in the injected eccentricity value. For more highly eccentric systems, $\chi_p$ posteriors show higher values when recovered with waveform models that neglect eccentricity, with a generally increasing trend that sees some variation due to the value of mean anomaly. When the injection is both eccentric and spin-precessing, eccentric analyses tend to recover the injected eccentricity values, while precessing posteriors are biased.

The paper is organised as follows. In Section \ref{sec:method}, we give an overview of the novel NR simulations used for injections and the Bayesian inference techniques used.  We present the results for non-spinning eccentric binaries in Section \ref{subsec:non-spin}, aligned-spin eccentric binaries in Section \ref{subsec:align-spin}, and precessing-spin eccentric binaries in Section \ref{subsec:prec-spin}. We conclude with a discussion in Section \ref{sec:concl}.

\section{Methodology}
\label{sec:method}

\subsection{Numerical Relativity Hybrids}
\label{subsec:NR_hyb}
For carrying out non-spinning, eccentric injections, we choose hybridized NR simulations, constructed in Chattaraj et al.~\cite{Chattaraj:2022tay}. These hybrid NR simulations are constructed by starting with eccentric, non-spinning NR simulations available in the public SXS catalog \citep{Scheel:2025jct} and hybridized with post-Newtonian inspiral waveforms~\cite{Boetzel:2019nfw, Ebersold:2019kdc, Tanay:2016zog, Moore:2016qxz}\footnote{It should be noted that the hybrids described in Chattaraj et al.~\citep{Chattaraj:2022tay} are not identical to the hybrids used in this study due to a difference in the PN prescription.}. The hybrids are labeled by an ID, which is the SXS simulation ID for the corresponding NR simulation used in the construction of the hybrids. The details of hybrids used in this study are given in Table~\ref{table:hybrids}.

\subsection{Numerical Relativity Simulations}
\label{subsec:NR_sims}

We run eccentric and/or precessing simulations of binary black holes using numerical relativity.
The simulations were performed using the Spectral Einstein Code (\texttt{SpEC}) developed by the Simulating eXtreme Spacetimes (SXS) collaboration \citep{SpECwebsite}. \texttt{SpEC} employs a multi-domain spectral discretization \citep{Kidder:1999fv, Scheel:2008rj, Szilagyi:2009qz, Hemberger:2012jz} to solve a first-order representation of the generalized harmonic system \citep{Lindblom:2005qh}. Excision surfaces are placed within apparent horizons \citep{Scheel:2008rj, Szilagyi:2009qz, Hemberger:2012jz, Ossokine:2013zga}, and constraint-preserving boundary conditions are used for the outer boundaries \cite{Lindblom:2005qh, Rinne:2006vv, Rinne:2007ui}. Superposed Kerr-Schild initial data \citep{Lovelace:2008hd} is constructed using \texttt{Spells} \citep{Pfeiffer:2002wt, Ossokine:2015yla}, which solves the extended conformal-thin sandwich equations \citep{York:1998hy, Pfeiffer:2002iy, Cook:2004kt}. The waveforms are post-processed, wherein they are extrapolated to infinity using \texttt{scri} \citep{scri} and corrected for the centre-of-mass drift \citep{Woodford:2019com}. The \texttt{SpEC} code was also optimized to make the most of the available hardware, enabling running simulations for longer, more efficiently. 
For the purpose of this study, we make use of a subset (see Table~\ref{table:ICTS_sims}) of these that fit our requirements. Specifically, we choose simulations such that the set includes diverse spin magnitudes and orientations while restricting to relatively low mass ratios ($q<6$) so as to remain in the parameter space where the analysis waveform \textsc{IMRESIGMA} is validated. We simulate at two different resolutions for the purpose of testing convergence. More details on the numerical simulations will be presented elsewhere \cite{eccprecictssims}. 

We use these NR simulations to inject aligned-spin and precessing-spin eccentric signals with different parameters. Additionally, we use the waveform \textsc{TEOBResumS-DALI}~\cite{Nagar:2018zoe} to inject aligned-spin eccentric signals with uniformly increasing values of eccentricity to study the trends in the source parameters.

\subsection{Bayesian inference}

\begin{table*}[t]
\setlength{\tabcolsep}{5.5pt}
\input{latex_sim_table_hybrids_twocol}

\caption{List of non-spinning, eccentric NR hybrid simulations used in this study. Columns include a unique hybrid/NR simulation ID for each simulation used in constructing the hybrids, total number of orbits (N\textsubscript{orbs}) after hybridization, mass ratio ($q=m_1/m_2$), eccentricity ($e_{20}$) and the mean anomaly ($l_{20}$) at a reference frequency of 20 Hz. See Sec.~\ref{subsec:NR_hyb} for further details.} 
\label{table:hybrids}
\end{table*}

In this work, we use Bayesian inference and stochastic sampling techniques to estimate the posterior distributions of the source parameters of the injected signals. The Bayesian posterior probability for a parameter $\Vec{\theta}$, given the data $\Vec{s}$ and a GW model $h$, is given by
\begin{equation}
  p(\Vec{\theta}|\Vec{s},h) = \frac{p(\Vec{s}|\Vec{\theta},h) p(\Vec{\theta},h)}{p(\Vec{s})}\,,
\end{equation}
where $p(\Vec{s}|\Vec{\theta},h)$ represents the likelihood, $p(\Vec{\theta})$ is the prior, and $p(\Vec{s}|h)$ represents the evidence. Further, Bayes factors, which can be calculated between recoveries with eccentric and quasi-circular models, are defined as:
\begin{equation}
    \mathcal{B}_\text{E/C} = \frac{p(\Vec{s}|h_E)}{p(\Vec{s}|h_C)}
\end{equation}
where $E$ and $C$ correspond to eccentric and quasi-circular recoveries respectively, and $h_i$ enumerates the waveform approximants under consideration.\footnote{For more information about the method, see Ref. \cite{Biwer:2018osg}.} 

\subsection{Parameter Estimation}
To estimate parameters, we use the \texttt{PyCBC Inference Toolkit}~\cite{Biwer:2018osg} and \texttt{bilby}~\cite{Ashton:2018jfp}, and explore the parameter space that includes chirp mass ($\mathcal{M}$), mass ratio ($q$), luminosity distance ($d_L$), inclination angle ($\iota$), time of coalescence ($t_c$), phase of coalescence ($\phi_c$), right ascension ($\alpha$), declination ($\delta$), and polarization angle ($\psi$). For aligned spin recoveries, we use two additional parameters corresponding to the $z$-components of the spin vectors \textit{viz.}~$\left({\chi_{\rm 1z}}~\&~{\chi_{\rm 2z}}\right)$. For recoveries with spin-precession, we use isotropic spin distribution sampling the six spin components in spherical polar coordinates \textit{viz.} the spin magnitudes ($a_i$) and the spin angles ($S_i^\Theta$, $S_i^\Phi$).\footnote{where $i=[1,2]$ corresponds to the binary components, and $\Theta$ and $\Phi$ indicate the polar and azimuthal angles respectively used in spherical polar coordinate system.} For recoveries with spins, we also obtain posteriors on two additional spin parameters. The first is the effective spin parameter, $\chi_\text{eff}$. This parameter captures the spin effects along the direction of the angular momentum axis and is defined as \cite{Ajith:2009bn, Santamaria:2010yb}:
\begin{equation}
    \chi_\text{eff} = \frac{m_1 \chi_\text{1z} + m_2 \chi_\text{2z}}{m_1 + m_2}, 
\label{eq:chi_eff}
\end{equation}
where $\chi_\text{1z}$ and $\chi_\text{2z}$ are the components of the two spin vectors in the direction of the angular momentum vector. The other parameter is the spin-precession parameter, $\rm{\chi_p}$, that measures the spin effects in-plane with the orbit of the binary, and is defined in terms of the perpendicular spin vectors, $S_{i\perp}=|\hat{L}\times (\vec{S_i}\times \hat{L})|$, where $\Vec{S}_i$ is the individual spin angular momentum vector of the compact object in the binary with mass $m_i$, and $\hat{L}$ represents the unit vector along the angular momentum axis of the binary. The effective spin-precession parameter can be written as \cite{Schmidt:2012rh, Hannam:2013oca, Schmidt:2014iyl}:
\begin{equation}
    {\chi_\mathrm{p}}=\frac{1}{A_{1}m_{1}^2} \max(A_{1}S_{1\perp}, A_{2} S_{2\perp}),
\label{eq:chi_p}
\end{equation}
where, $A_{1}=2+(3/2q)$ and $A_{2}=2+(3q/2)$ are mass parameters defined in terms of the mass ratio $q=m_1/m_2>1$.

For eccentric recoveries with \textsc{TaylorF2Ecc}~\cite{Moore:2016qxz, Kim:2019abc}, we sample on the eccentricity parameter ($e$), and for eccentric recovery with \textsc{IMRESIGMA}~\citep{Paul:2024ujx}, we sample over eccentricity ($e$) as well as the mean anomaly ($l$) parameter in addition to the standard parameters for aligned-spin recovery. The complete information on the prior models and ranges, along with the sampler settings used for the analysis is included in Table \ref{table:priors} in Appendix \ref{appendix:priors}.

\begin{figure*}[th!]
    \centering
    \includegraphics[width=\linewidth]{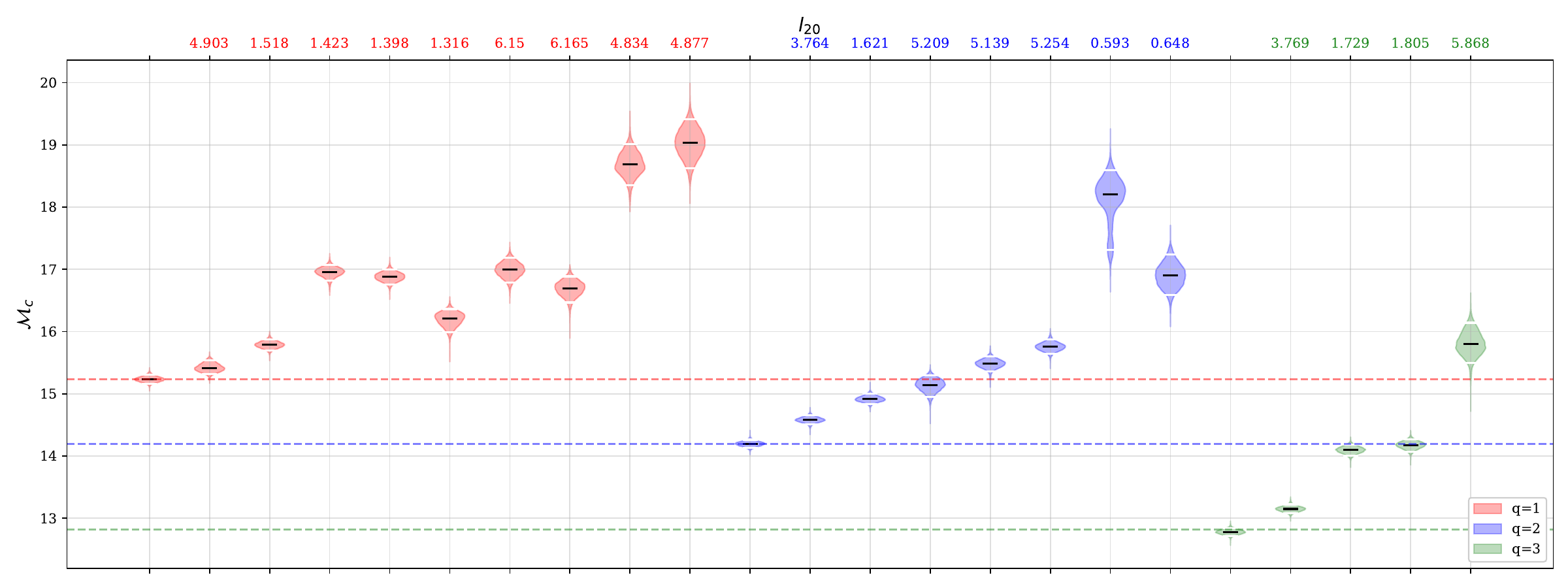}
    
    \includegraphics[width=\linewidth]{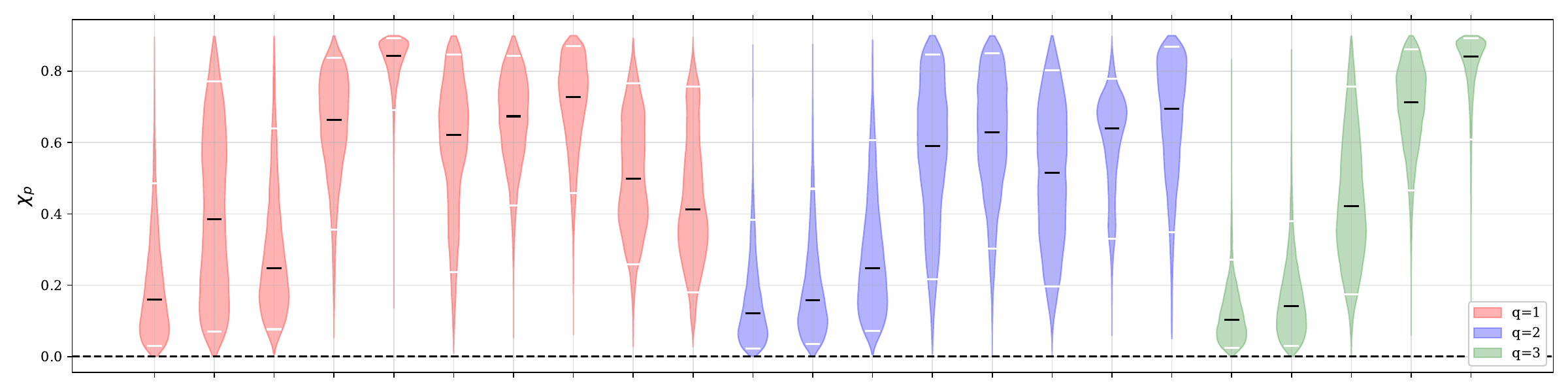}

    \includegraphics[width=\linewidth]{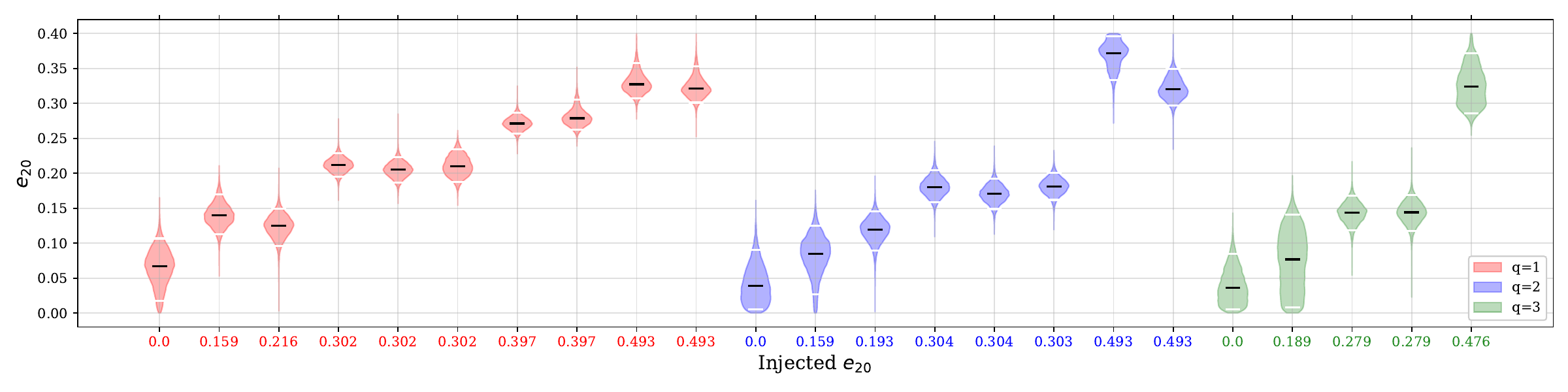}
    \caption{The recovery of various parameters of non-spinning hybrid injections (see Sec.~\ref {subsec:NR_hyb}). The colours red, blue, and green indicate mass ratios 1, 2, and 3, respectively. The horizontal dashed lines in the top two plots denote injected values for the respective parameters. The small black lines inside the violins mark the median values of the posteriors, and the small white lines denote the 90\% credible intervals. Top: Recovery of chirp mass ($\mathcal{M}_c$) parameter with \textsc{IMRPhenomXP}~\citep{Pratten:2020ceb} waveform in precessing-spin configuration. Middle: Recovery of precessing spin parameter ($\chi_p$) when the injections are recovered with \textsc{IMRPhenomXP} in precessing-spin configuration. Bottom: Recovery of eccentricity ($e_{20}$) parameter at a reference frequency of 20 Hz when the injected signals are recovered with \textsc{TaylorF2Ecc}~\cite{Moore:2016qxz, Kim:2019abc} in non-spinning eccentric configuration. The bottom and top labels on the horizontal axis list the injected values of eccentricity ($e_{20}$) and mean anomaly ($l_{20}$).}
    \label{fig:chip_e20_hybrids}
\end{figure*}

We use the HLV network \citep{LIGOScientific:2014pky, VIRGO:2014yos} with design sensitivities of Advanced LIGO \cite{PyCBC-PSD:aLIGO} and Virgo \cite{PyCBC-PSD:AdvVirgo} detectors to perform all the parameter estimation analyses shown here. All injections performed in this work include only the dominant modes ($\ell=2, |m|=2$) of gravitational radiation and are performed in zero-noise. The injections are created for BBH mergers at a luminosity distance of 410 Mpc, inclined at an angle of $30^\circ$ with the line-of-sight. We have arbitrarily chosen the $\alpha$, $\delta$, and $\psi$ angles to be $164^\circ$, $60^\circ$, and $60^\circ$ respectively. The geocent time was set to $1137283217s$. 

We follow the approach of computing eccentricities from the waveform at a quadrupolar mode reference frequency of $20$~Hz. To calculate the eccentricity and mean anomaly values, we use the package \texttt{gw\_eccentricity}~\cite{Shaikh:2023ypz, Shaikh:2025tae}, which estimates the eccentricity values using the frequency evolution of the waveform. 
\footnote{See also \cite{Islam:2025oiv}, an alternative framework for measuring eccentricity directly from the waveform.} For systems with 0 eccentricity, the value of mean anomaly is not defined; hence, the blank fields in Table \ref{table:hybrids}.

\section{Results}
\label{sec:results}

\subsection{Eccentric and non-spinning binaries}
\label{subsec:non-spin}

We inject non-spinning, quasi-circular as well as eccentric signals using IMR hybrids constructed in Ref.~\citep{Chattaraj:2022tay} and described in Sec.~\ref{subsec:NR_hyb}. They have mass ratios of $q=(1, 2, 3$). In addition, we also use a quasi-circular SXS simulation (\textsc{SXS:BBH:1132}). We set the total mass of the system to $M=35$~M$_\odot$. Details of the simulations used in this study, including their eccentricities and mean anomaly values at the reference frequency of $20$~Hz, are shown in Table \ref{table:hybrids}.

First, we explore the biases introduced in the source parameters recovered via parameter estimation (PE) of GW events when eccentricity is ignored, i.e.~we inject an eccentric signal but do not use eccentric waveforms for recovery. For this exercise, we use the phenomenological waveform model \textsc{IMRPhenomXP}~\citep{Pratten:2020ceb}. We then analyse the signals with an eccentric waveform model \textsc{TaylorF2Ecc} (which uses the same PN prescription as \textsc{IMRPhenomXP}) to get bounds on the eccentricity parameter. Since the signals simulated here are low mass ($M = 35 M_\odot$), they are inspiral dominated and an inspiral model such as \textsc{TaylorF2Ecc} should suffice.

In the top two panels of Fig.~\ref{fig:chip_e20_hybrids}, we show the recovered marginal distributions of chirp mass ($\mathcal{M}_c$) and effective spin-precession ($\chi_p$) when the hybrid injections are analysed with a spin-precessing quasi-circular waveform model. We observe that the bias in $\mathcal{M}_c$ generally increases with the value of the injected eccentricity, although there are small variations seen due to a dependence on the mean anomaly values.\footnote{In the following section, we fix the mean anomaly values for the injections and see that the variation in the trends is quite reduced.} We also find that for low values of eccentricity, the $\chi_p$ posteriors are either peaking near zero or returning the prior distribution (uninformative). However, for high eccentricities, the $\chi_p$ posteriors peak at higher values. This indicates that the recovery of an eccentric non-spinning signal with a non-eccentric precessing model may result in a false indication of precession for sufficiently high values of eccentricity.

In the bottom panel of Fig.~\ref{fig:chip_e20_hybrids}, we show the recovered marginal distributions of the eccentricity parameter as a function of the injected value of eccentricity.

Further, we calculate Bayes factors in favor of the eccentric model over the circular, and find that, on average, as the eccentricity increases, the Bayes factors ($\mathcal{B}_{E/C}$) increase, with variations due to different mean anomaly values. This comparison of Bayes factors is done in a more organised fashion in the next section for aligned spin eccentric injections, where the injected mean anomaly values are fixed.


\subsection{Eccentric and spin-aligned binaries}
\label{subsec:align-spin}


\begin{figure}
    \centering
    \includegraphics[width=\linewidth]{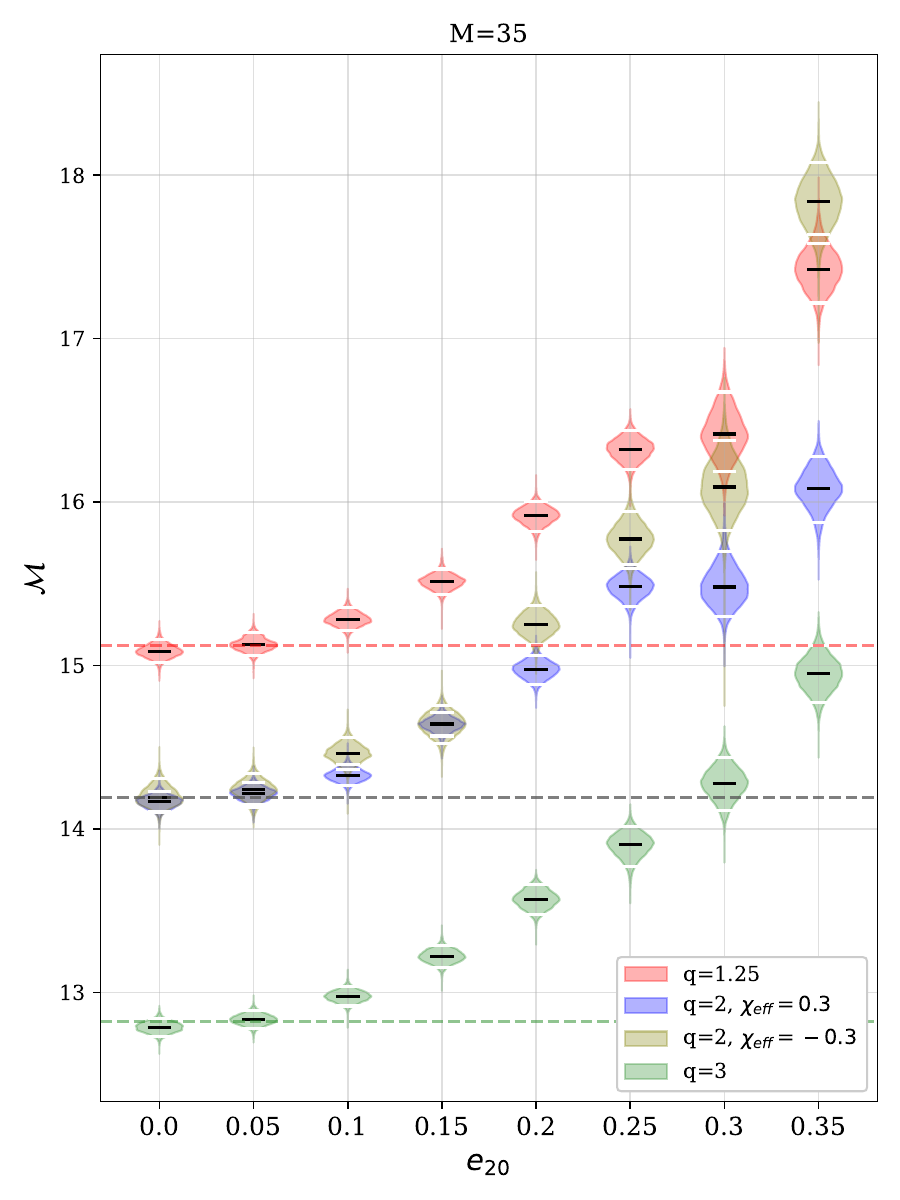}
    \caption{Recovery of the chirp mass ($\mathcal{M}_c$) parameter when injections are aligned-spin eccentric signals, generated using the \textsc{TEOBResumS-DALI} waveform model. The colours red, blue, and green indicate mass ratios $1.25, 2$, and $3$, respectively for aligned-spin injections whereas the anti-aligned spin case for $q=2$ is denoted in olive. The horizontal dashed lines denote injected values for the respective mass ratio cases. The small black lines inside the violins mark the median values of the posteriors, and the small white lines denote the 90\% credible intervals. The injected signals are recovered with the \textsc{IMRPhenomXP} waveform model in the precessing-spin configuration.}
    \label{fig:Mc_TEOB}
\end{figure}

\begin{figure}[h]
    \centering
    \includegraphics[width=\linewidth]{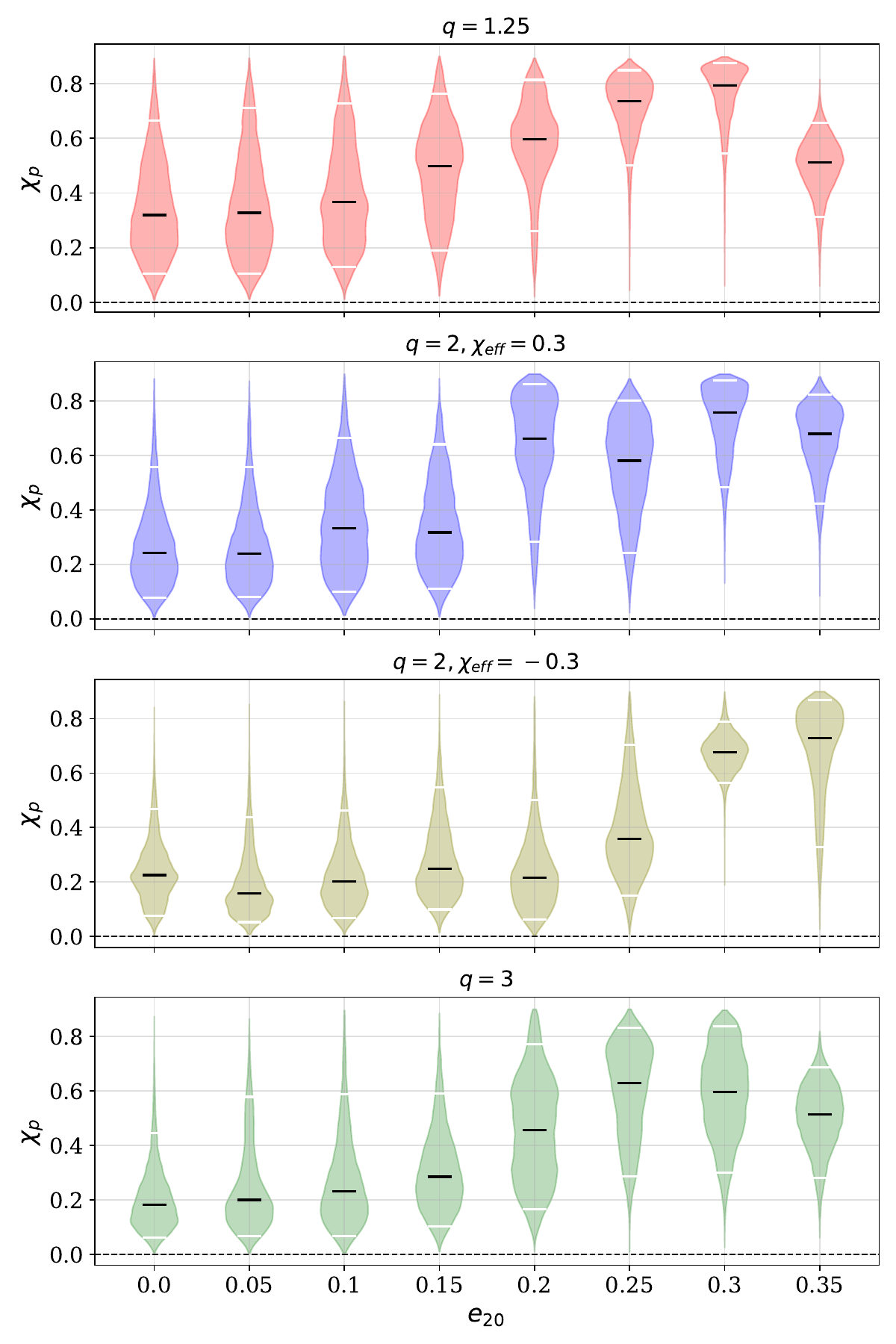}
    \caption{Recovery of spin-precession parameter ($\chi_p$) when injections are aligned-spin eccentric signals generated using the \textsc{TEOBResumS-DALI} waveform model. The colours red, blue, and green indicate mass ratios 1.25, 2, and 3, respectively for aligned-spin injections whereas the anti-aligned spin cases for $q=2$ are denoted in olive. The horizontal dashed line denotes the injected value. The small black lines inside the violins mark the median values of the posteriors, and the small white lines denote the 90\% credible intervals. The injections are recovered with \textsc{IMRPhenomXP} in the precessing-spin configuration.}
    \label{fig:chip_TEOB}
\end{figure}

\begin{figure}[h]
    \centering
    \includegraphics[width=\linewidth]{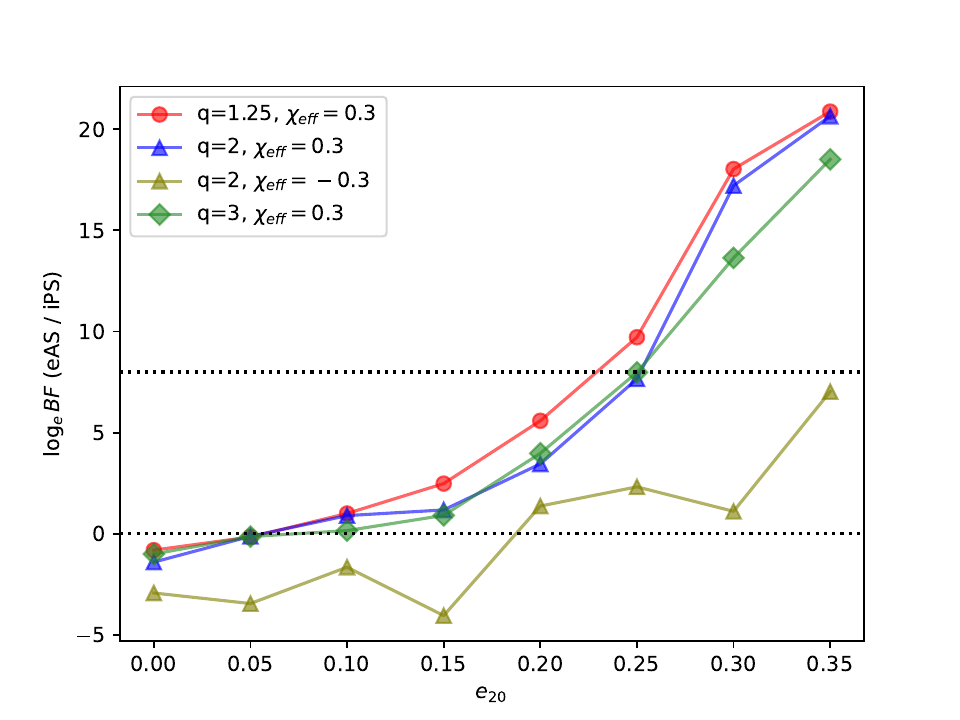}
    \caption{Aligned spin injections with \textsc{TEOBResumS-DALI}. Bayes factors for the eccentric aligned-spin recovery with \textsc{TaylorF2Ecc} over the quasi-circular precessing-spin recovery with \textsc{IMRPhenomXP}. The  \textsc{IMRPhenomXP} analysis is truncated at 110 Hz to facilitate a fair comparison in the calculation of the Bayes factors against \textsc{TaylorF2Ecc}.}
    \label{fig:bayes_factors_TEOB}
\end{figure}

We inject aligned-spin eccentric signals using the waveform model \textsc{TEOBResumS-DALI} \cite{Nagar:2018zoe}, and recover them using \textsc{IMRPhenomXP} in quasi-circular, precessing configuration, and with \textsc{TaylorF2Ecc} in the aligned-spin eccentric configuration. As in the previous case, the total mass is taken to be $35$ M$_\odot$, and here we choose to inject signals with mass ratios $q=1.25, 2, 3$. The injected spin magnitudes are $\chi_\text{1z} = \chi_\text{2z} = 0.3 = \chi_\text{eff}$ (refer Eq.~\eqref{eq:chi_eff}). We also explore the anti-aligned spin case ($\chi_\text{1z} = \chi_\text{2z} = -0.3 = \chi_\text{eff}$) for $q=2$. We vary the eccentricity from 0 to 0.35 in steps of 0.05 to analyze the behaviour of the recovery posteriors, when all the other parameters are kept the same across injections, for a given mass ratio. 

Fig.~\ref{fig:Mc_TEOB} shows the bias in the chirp mass posteriors when an eccentric signal is recovered with a quasi-circular waveform in the precessing-spin configuration. The bias increases as a function of increasing eccentricity, following the same general trend as seen in the non-spinning injection cases (Sec.~\ref{subsec:non-spin}). Additionally, as the injected mean anomaly value is the same for all signals here, there are no significant variations in the observed bias as opposed to the non-spinning case. We also observe in Fig.~\ref{fig:chip_TEOB} that even though the injections are aligned-spin, the $\chi_p$ posteriors peak at higher values at large eccentricities. This indicates that a quasi-circular precessing-spin waveform is unable to reliably recover the true value of the spin-precession parameter (0 in this case) when the injected signals have high eccentricity. 

Furthermore, we observe that the trends remain roughly the same as the mass-ratio varies. However, this may change when higher modes are included in injections and/or recovery of the eccentric signals. In this work, we have included only the ($\ell=2, |m|=2$) modes, and we leave the analysis of eccentric signals with higher modes for future work.

Further, we pick three aligned-spin systems (tabulated in Table~\ref{table:ICTS_sims}) from the local catalog of Numerical simulations discussed in Sec.~\ref{subsec:NR_sims}, inject them, and recover with the quasi-circular precessing model. We find the same result as before and are denoted by the first three cases in Fig.~\ref{fig:chip_e20_ICTS} (left panel). A detailed discussion on the results of NR simulations is included in the following section. 

\begin{table*}[ht]
\setlength{\tabcolsep}{5pt}
\input{latex_sim_table_ICTS_sims}

\caption{Spinning, eccentric NR simulations in increasing order of eccentricity from the ICTS catalog (Sec.~\ref{subsec:NR_sims}) used in the work. Columns include a unique NR simulation ID for each simulation, number of orbits (N\textsubscript{orbs}), mass ratio ($q=m_1/m_2$), total mass ($M$) injected, reference frequency ($f_\text{ref}$) at which eccentricity and mean anomaly are defined, eccentricity ($e_\text{ref}$), mean anomaly ($l_\text{ref}$), effective spin parameter ($\chi_\text{eff}$) defined in Eq.~\eqref{eq:chi_eff}, and spin-precession parameter ($\chi_p$) defined in Eq.~\eqref{eq:chi_p}. Eccentricities $e_\text{ref}$ and mean anomalies $l_\text{ref}$ were estimated from the waveforms using the \texttt{gw\_eccentricity} package.}
\label{table:ICTS_sims}
\end{table*}

Since an analysis with a quasi-circular precessing-spin waveform yields high values of $\chi_p$, it is natural to analyze these signals with an eccentric and aligned-spin waveform, and compute the corresponding Bayes factors. The Bayes factors, for aligned-spin and anti-aligned spin injections with \textsc{TEOBResumS-DALI}, comparing the eccentric aligned-spin recovery with quasi-circular precessing-spin recovery are shown in Fig.~\ref{fig:bayes_factors_TEOB}. We find that for high values of eccentricity, the eccentric and aligned-spin model is considerably preferred over the quasi-circular precessing-spin model. This is in agreement with the conclusions drawn in Romero-Shaw et al \citep{Romero-Shaw:2022fbf} where Bayes factors prefer an eccentric model for signals with a sufficient number of cycles in the band. It can be seen that for the anti-aligned case (shown in olive in Fig.~\ref{fig:bayes_factors_TEOB}), while the general trend is consistent with the aligned-spin cases, the Bayes factors are lower. This may be due to the differences in waveform modelling where \textsc{IMRPhenomXP}, in general, performs better than \textsc{TaylorF2Ecc} for negative spins since it is better calibrated in that region of parameter space. Since \textsc{TaylorF2Ecc} is an inspiral-only waveform, for the computation of the Bayes factor, we truncate the likelihood calculation at $110$~Hz for both models in accordance with the choice of total mass as described above.


\subsection{Eccentric and spin-precessing binaries}
\label{subsec:prec-spin}

\begin{figure*}
    \centering
    \includegraphics[width=0.49\linewidth]{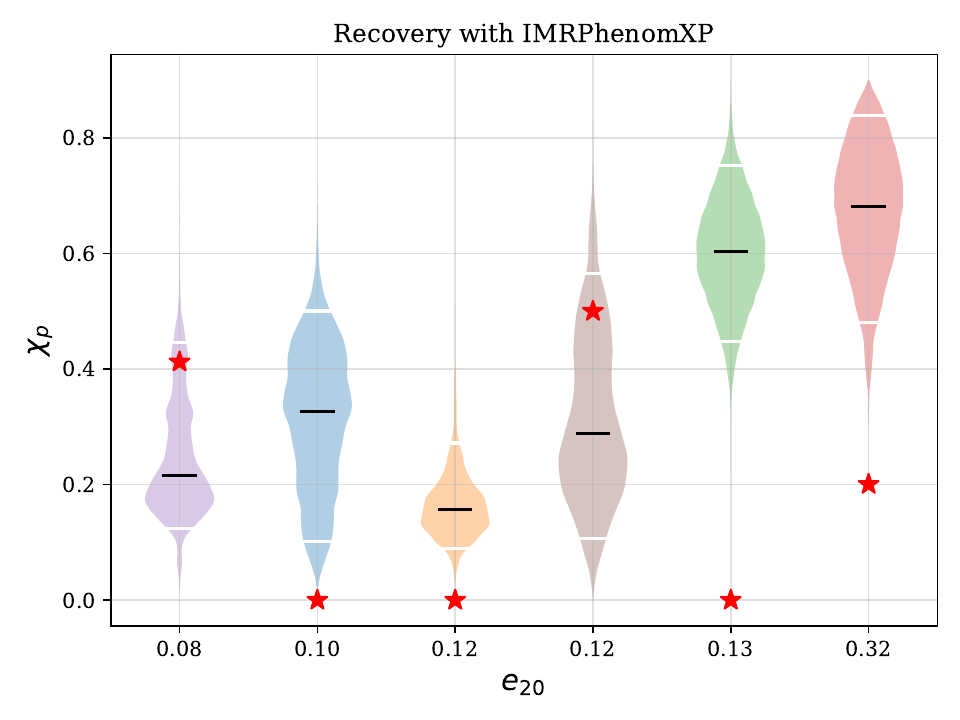}
    \includegraphics[width=0.49\linewidth]{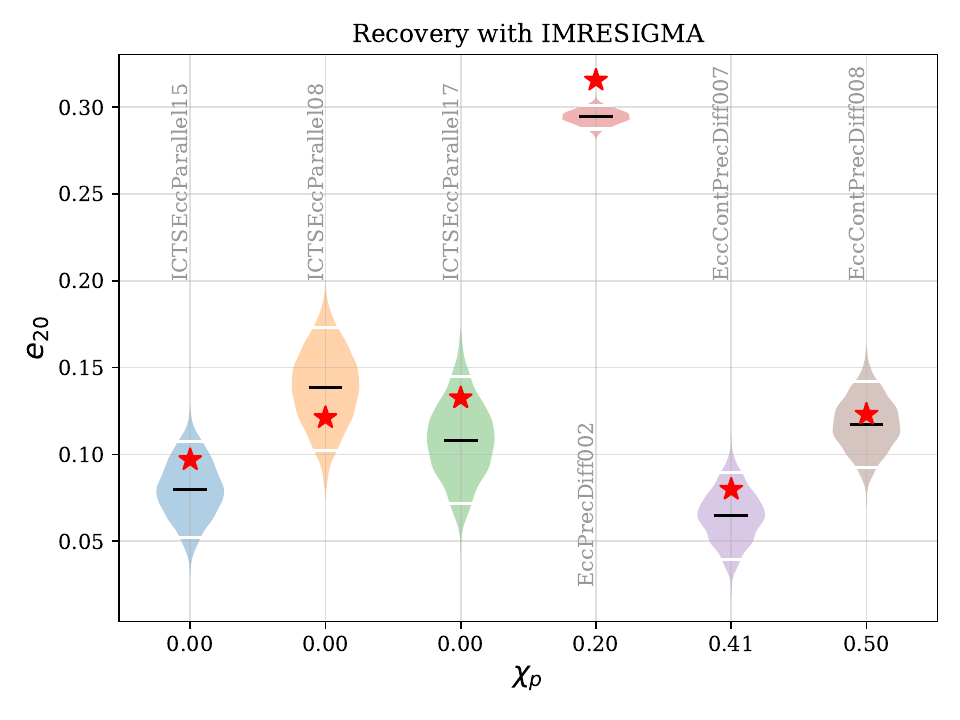}
    \caption{Figures show the recovery of the spin-precession parameter ($\chi_p$) and eccentricity ($e_{20}$) when the injections are using the NR simulations described in Sec.~\ref{subsec:NR_sims} and Table~\ref{table:ICTS_sims}. The small black lines inside the violins mark the median values of the posteriors, and the small white lines denote the 90\% credible intervals. Left: Recovery of the spin-precession parameter ($\chi_p$) when the injections are recovered with \textsc{IMRPhenomXP} in quasi-circular precessing-spin configuration. The posteriors are arranged in ascending order of injected eccentricities ($e_{20}$) at a reference frequency of $20$ Hz calculated using the \texttt{gw\_eccentricity} package. The injected values of $\chi_p$ are shown as red stars on the plot. Right: Recovery of the eccentricity parameter ($e_{20}$) at a reference frequency of $20$Hz when the injection signals are analysed with \textsc{IMRESIGMA} in aligned-spin eccentric configuration. The posteriors are arranged in ascending order of injected $\chi_p$ values. The red stars denote the injected value of the eccentricities ($e_{20}$). For clarity, we have assigned different colours to different simulations, and these are consistent across both panels. As the definition of eccentricity is different in \textsc{IMRESIGMA} waveform as opposed to the value calculated using the \texttt{gw\_eccentricity} package for the injections, the posteriors for eccentricity have been converted to the values obtained from the \texttt{gw\_eccentricity} package for \textsc{IMRESIGMA} waveform for a fair comparison with the injected values.}
    \label{fig:chip_e20_ICTS}
\end{figure*}

\begin{figure}[ht!]
    \centering
    \includegraphics[width=\linewidth]{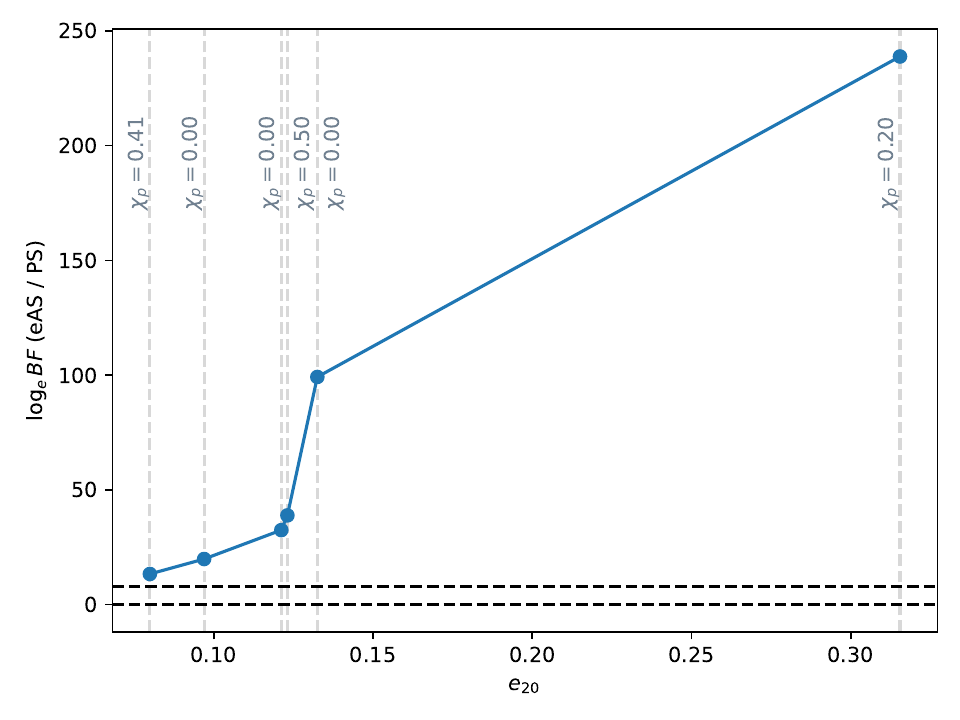}
    \caption{Bayes factors for eccentric aligned-spin recovery (eAS) over quasi-circular precessing-spin recovery (PS) when the injection signals are created using the NR simulations described in Sec.~\ref{subsec:NR_sims} and Table~\ref{table:ICTS_sims}. Both are IMR recoveries where \textsc{IMRESIGMA} is used for the former, whereas \textsc{IMRPhenomXP} is used for the latter. The $\chi_p$ values corresponding to the simulations at each eccentricity are mentioned for each point on the plot.}
    \label{fig:bayes_factors_ICTS}
\end{figure}

We inject eccentric precessing-spin signals using NR simulations run with the \texttt{SpEC} code, detailed in the Sec.~\ref{subsec:NR_sims}. Due to the varying length of the simulations, we have used different total masses for the injections to include all the cycles. Also, different simulations have different values of eccentricity, mean anomaly, mass ratios, and spins. The details of injection parameters for various simulations used are given in Table~\ref{table:ICTS_sims}. We analyze these signals in precessing-spin configuration using the quasi-circular waveform \textsc{IMRPhenomXP}. As many of these signals are short in duration, for eccentric recoveries, we use the IMR eccentric, aligned-spin model \textsc{IMRESIGMA} for the analysis, instead of \textsc{TaylorF2Ecc}. 

Figure~\ref{fig:chip_e20_ICTS} shows recovery of the spin-precession parameter ($\chi_p$) and eccentricity ($e_{20}$) at a reference frequency of 20 Hz. On the left, we show the recovery of $\chi_p$ when the injections are recovered with \textsc{IMRPhenomXP} in a quasi-circular precessing-spin configuration. The posteriors are arranged in ascending order of injected eccentricity ($e_{20}$) at a reference frequency of 20 Hz, calculated using the \texttt{gw\_eccentricity} package. For comparison, the injected values of $\chi_p$ are shown as red stars on the plot. In the figure on the right, we show the recovery of eccentricity parameter ($e_{20}$) when the injection signals are analysed with \textsc{IMRESIGMA} in aligned-spin eccentric configuration. These posteriors are arranged in ascending order of injected $\chi_p$ values. The red stars denote the injected value of eccentricity ($e_{20}$) for reference. Since the definition of eccentricity is different in \textsc{IMRESIGMA} waveform as compared to the value calculated using the \texttt{gw\_eccentricity} package for the injections, the eccentricity posteriors have been converted to the values computed from the \texttt{gw\_eccentricity} package.

It can be seen that with an increase in injected eccentricity, the posteriors on the $\chi_p$ parameter are consistently overestimated, and for high eccentricity values, the recovered $\chi_p$ posteriors exclude the injected values within 90\% credible intervals. On the other hand, with an increase in the $\chi_p$ values, the eccentricity posteriors are not significantly affected. Even for $\chi_p$ of 0.5, the aligned-spin eccentric waveform is able to recover the injected value of eccentricity within 90\% credible interval. A further calculation of Bayes factors (Fig.~\ref{fig:bayes_factors_ICTS}) confirms that with an increasing value of eccentricity, an eccentric aligned-spin waveform model (\textsc{IMRESIGMA}) is consistently preferred over a quasi-circular precessing-spin waveform model (\textsc{IMRPhenomXP})\footnote{We note that due to differences in modelling approach between the two waveform models, there may be some systematics which affect the Bayes factor values, but we believe the overall trend would not be affected considerably since the injections have been made using NR simulations.}.

\section{Conclusions} 
\label{sec:concl}

We have performed parameter estimation of eccentric signals with varying degrees of spin complexity to study the effect of spins and eccentricity on the source parameters.\footnote{The parameter biases observed in these results are dependent on SNRs, which, in this study, fall in the range of typical SNRs observed in the GW event catalogs.} We start with non-spinning signals in Sec .~\ref {subsec:non-spin}, where we inject hybrids and recover with a quasi-circular precessing waveform. We observe that for high values of injected eccentricity, the posterior on the spin-precessing parameter ($\chi_p$) deviates away from the injected value of zero. This indicates that at high eccentricities, a quasi-circular precessing waveform is unable to distinguish eccentricity from precession effects, and such signals should be further investigated by analysing with eccentric waveform models. We also see biases in the chirp mass parameter, which is similar to the biases obtained in Divyajyoti et al.\cite{Divyajyoti:2023rht} and Das et al. \citep{Das:2024zib}. In these works, the hybrid injections were recovered with quasi-circular non-spinning waveform.\footnote{In Das et al. \cite{Das:2024zib}. The injections were for a 40~$M_\odot$ system, whereas in this paper, we use 35~$M_\odot$ systems, although the general trend for the biases remains the same.} Slight variations are seen in the general trend for biases, which can be attributed to the differences in the mean anomaly values. 

To mitigate the effect of mean anomalies, in the following section (Sec.~\ref{subsec:align-spin}), we inject aligned-spin signals using the waveform \textsc{TEOBResumS-DALI}, with the same mean anomaly values. In this case, we observe clear trends in the bias of chirp mass parameters when these aligned-spin eccentric signals are analyzed with a quasi-circular precessing-spin waveform. Moreover, we observe that with an increase in the injected eccentricity value, the posteriors on $\chi_p$ peak away from zero, again indicating that a quasi-circular precessing-spin waveform may not be sufficient to recover the correct spins of eccentric signals. Although a further investigation into these signals reveals that when Bayes factors are calculated between an eccentric aligned-spin recovery and a quasi-circular precessing-spin recovery, the eccentric model is clearly preferred over a quasi-circular model, and the Bayes factors increase with eccentricity. 

Finally, in Sec .~\ref {subsec:prec-spin}, we analyze a few NR simulations of systems that are both spin-precessing and eccentric. Due to the varying lengths and other properties of these simulations, the injected signals differ in total mass, mass ratio, spin values, eccentricities, and mean anomaly. Nevertheless, we observe interesting trends. When these eccentric, spin-precessing signals are analyzed with a quasi-circular precessing waveform, the recovered posteriors on $\chi_p$ deviate from the injected values. This deviation is low at low values of eccentricities, but becomes significant as eccentricity increases. On the other hand, when the same signals are analysed with an eccentric, aligned-spin waveform, the recovered eccentricity posteriors are not significantly biased as the injected $\chi_p$ increases. This suggests that, when an eccentric and spin-precessing system is detected via its GW emission, eccentricity estimates obtained through Bayesian inference that neglects spin-precession are not drastically affected, while spin-precession inferences that neglect eccentricity can be significantly biased. Furthermore, the Bayes factors calculated between these two recoveries show a clear preference for the eccentric, aligned-spin waveform over the quasi-circular precessing-spin waveform, and the Bayes factors increase with an increase in the injected eccentricity values. 

Our results demonstrate the complex interplay between the effects of eccentricity and spin-precession in GW signals when either of the two effects is ignored in parameter estimation. It underlies the importance of a thorough investigation of GW signals which show high spin-precession. It also highlights the need for comprehensive IMR eccentric spin-precessing waveforms, which need to be employed in the parameter estimation of signals that exhibit signs of eccentricity and/or spin-precession in order to mitigate bias in inferred parameters of the sources.

\acknowledgments

We thank Nihar Gupte for the useful comments on the manuscript. 
We thank Pratul Manna for helping us with the calculation of eccentricity values using the \texttt{gw\_eccentricity} package for hybrid injections. 
D.J. acknowledges the Science and Technology Facilities Council (STFC) for support through grants ST/V005618/1 and ST/Y004272/1. 
I.M.R-S. acknowledges support received from the Ernest Rutherford Fellowship of the Science and Technology Facilities Council, grant number UKRI2423. V.P's work was supported by the Department of Atomic Energy, Government of India, under Project No. RTI4001, and the National Science Foundation awards PHY-2309064 and PHYS-2308886.  C.K.M. acknowledges the support of ANRF’s Core Research Grant No.\ CRG/2022/007959. Computations were performed on the \texttt{powehi} workstation in the Department of Physics, IIT Madras, and CIT cluster provided by the LIGO Laboratory. Numerical relativity simulations and some parameter estimation runs were performed on the \texttt{sonic} HPC at ICTS-TIFR. The authors are grateful for computational resources provided by the LIGO Laboratory and supported by National Science Foundation Grants Np.~PHY-0757058 and No.~PHY-0823459. We used the following software packages: {\tt LALSuite}~\cite{lalsuite}, {\tt PyCBC}~\cite{alex_nitz_2020_4134752}, {\tt bilby}~\cite{Ashton:2018jfp}, {\tt NumPy}~\cite{Harris:2020xlr}, {\tt Matplotlib}~\cite{2007CSE.....9...90H}, {\tt Seaborn}~\cite{Waskom2021}, {\tt jupyter}~\cite{soton403913}, {\tt dynesty}~\cite{Speagle:2019ivv}, {\tt corner}~\cite{corner}.
This document has LIGO preprint number {\tt LIGO-P2500606}.

\appendix

\section{Priors and sampler settings used for parameter estimation}
\label{appendix:priors}

The priors on various parameters used for non-spinning, aligned-spin, and precessing-spin analyses are listed in Table \ref{table:priors}. For analysis with \texttt{PyCBC Inference toolkit}, we use the \texttt{dynesty} sampler with \texttt{rwalk} method, \texttt{nlive=2000}, \texttt{walks=200}, \texttt{nact=20}, and \texttt{dlogz=0.1}. In our analysis using \texttt{bilby}, we set \texttt{nlive=1000}, keeping all other sampler settings unchanged.

\begin{table}[ht]
\def\arraystretch{1.3}
\begin{tabular}{|c|c|c|}
\hline
\textbf{Parameter} & \textbf{Prior} & \textbf{Range} \\ \hline
$\mathcal{M}$ & \begin{tabular}[c]{@{}c@{}}Uniform in \\ component masses \end{tabular} & $5 \text{ - } 50~M_\odot$ \\ \hline
$q$ & \begin{tabular}[c]{@{}c@{}}Uniform in \\ component masses \end{tabular} & \begin{tabular}[c]{@{}c@{}} $1 \text{ - } 5$ \\ $1 \text{ - } 15$\footnote{For NR simulations listed in Table~\ref{table:ICTS_sims}} \end{tabular} \\ \hline
$d_L$ & Uniform radius & $100 \text{ - } 3000$ Mpc \\ \hline
$\iota$ & Uniform sine & $0 \text{ - } \pi$ \\ \hline
$t_c$ & Uniform & $t_\text{gps}+(-0.1 \text{ - } 0.1)$~s \\ \hline
$\phi_c$ & Uniform & $0 \text{ - } 2\pi$ \\ \hline
$\chi_{i\text{z}}$\footnote{\label{note:index}where $i=[1,2]$ refers to the binary components}\textsuperscript{, }\footnote{only used for aligned-spin recovery} & Uniform & $0 \text{ - } 0.9$ \\ \hline
$a_1$, $a_2$\footnote{\label{note:prec-spin}only used for precessing spin recovery} & Uniform & $0 \text{ - } 0.9$ \\ \hline
$(S_i^\Theta + S_i^\Phi)$\textsuperscript{\ref{note:index}, \ref{note:prec-spin}} & Uniform solid angle & \begin{tabular}[c]{@{}c@{}} $\Theta \in (0,\pi)$, \\ $\Phi \in (0,2\pi)$ \end{tabular} \\ \hline
$(\alpha + \delta)$ & Uniform sky & \begin{tabular}[c]{@{}c@{}} $\delta \in (\pi/2,-\pi/2)$, \\ $\alpha \in (0,2\pi)$ \end{tabular} \\ \hline
$\psi$ & Uniform & $0 \text{ - } 2\pi$ \\ \hline
$e$\footnote{\label{note:ecc}only used for eccentric recovery} & Uniform & $0 \text{ - } 0.4$ \\ \hline
$l$\textsuperscript{\ref{note:ecc}} & Uniform & $0 \text{ - } 2\pi$ \\ \hline
\end{tabular}
\caption{Priors for parameters in various quasi-circular and eccentric recoveries.}
\label{table:priors}
\end{table}



\bibliographystyle{apsrev4-1}
\bibliography{master_refs_new}
\end{document}

%% file: latex_sim_table_hybrids_twocol.tex
\begin{minipage}[t]{0.45\textwidth}
\centering
\begin{tabular}[t]{|c|c|c|c|c|c|}
\hline
\toprule
S.No & Simulation ID & N\textsubscript{orbs} & q & $e_{20}$ & $l_{20}$ \\
\hline
\midrule
1 	 & 	 SXS:BBH:1132 	 & 	 53 	 & 	 1 	 & 	 0.000 	 & 	 - \\ \hline
2 	 & 	 HYB:SXS:BBH:1167 	 & 	 48 	 & 	 2 	 & 	 0.000 	 & 	 - \\ \hline
3 	 & 	 HYB:SXS:BBH:1221 	 & 	 56 	 & 	 3 	 & 	 0.000 	 & 	 - \\ \hline
4 	 & 	 HYB:SXS:BBH:1355 	 & 	 41 	 & 	 1 	 & 	 0.159 	 & 	 4.903 \\ \hline
5 	 & 	 HYB:SXS:BBH:1356 	 & 	 39 	 & 	 1 	 & 	 0.216 	 & 	 1.518 \\ \hline
6 	 & 	 HYB:SXS:BBH:1357 	 & 	 36 	 & 	 1 	 & 	 0.302 	 & 	 1.423 \\ \hline
7 	 & 	 HYB:SXS:BBH:1358 	 & 	 35 	 & 	 1 	 & 	 0.302 	 & 	 1.398 \\ \hline
8 	 & 	 HYB:SXS:BBH:1359 	 & 	 36 	 & 	 1 	 & 	 0.302 	 & 	 1.316 \\ \hline
9 	 & 	 HYB:SXS:BBH:1360 	 & 	 31 	 & 	 1 	 & 	 0.397 	 & 	 6.150 \\ \hline
10 	 & 	 HYB:SXS:BBH:1361 	 & 	 31 	 & 	 1 	 & 	 0.397 	 & 	 6.165 \\ \hline
11 	 & 	 HYB:SXS:BBH:1362 	 & 	 25 	 & 	 1 	 & 	 0.493 	 & 	 4.834 \\ \hline
12 	 & 	 HYB:SXS:BBH:1363 	 & 	 25 	 & 	 1 	 & 	 0.493 	 & 	 4.877 \\ \hline
\bottomrule
\end{tabular}
\end{minipage}
\quad
\begin{minipage}[t]{0.45\textwidth}
\centering
\begin{tabular}[t]{|c|c|c|c|c|c|}
\hline
\toprule
S.No & Simulation ID & N\textsubscript{orbs} & q & $e_{20}$ & $l_{20}$ \\
\hline
\midrule
13 	 & 	 HYB:SXS:BBH:1364 	 & 	 46 	 & 	 2 	 & 	 0.159 	 & 	 3.764 \\ \hline
14 	 & 	 HYB:SXS:BBH:1365 	 & 	 44 	 & 	 2 	 & 	 0.193 	 & 	 1.621 \\ \hline
15 	 & 	 HYB:SXS:BBH:1366 	 & 	 39 	 & 	 2 	 & 	 0.304 	 & 	 5.209 \\ \hline
16 	 & 	 HYB:SXS:BBH:1367 	 & 	 40 	 & 	 2 	 & 	 0.304 	 & 	 5.139 \\ \hline
17 	 & 	 HYB:SXS:BBH:1368 	 & 	 40 	 & 	 2 	 & 	 0.303 	 & 	 5.254 \\ \hline
18 	 & 	 HYB:SXS:BBH:1369 	 & 	 28 	 & 	 2 	 & 	 0.493 	 & 	 0.593 \\ \hline
19 	 & 	 HYB:SXS:BBH:1370 	 & 	 28 	 & 	 2 	 & 	 0.493 	 & 	 0.648 \\ \hline
20 	 & 	 HYB:SXS:BBH:1371 	 & 	 52 	 & 	 3 	 & 	 0.189 	 & 	 3.769 \\ \hline
21 	 & 	 HYB:SXS:BBH:1372 	 & 	 48 	 & 	 3 	 & 	 0.279 	 & 	 1.729 \\ \hline
22 	 & 	 HYB:SXS:BBH:1373 	 & 	 48 	 & 	 3 	 & 	 0.279 	 & 	 1.805 \\ \hline
23 	 & 	 HYB:SXS:BBH:1374 	 & 	 35 	 & 	 3 	 & 	 0.476 	 & 	 5.868 \\ \hline
\bottomrule
\end{tabular}
\end{minipage}

%% file: latex_sim_table_ICTS_sims.tex
\begin{tabular}{|c|c|c|c|c|c|c|c|c|c|}
\hline
S.No & Simulation ID & N\textsubscript{orbs} & q & M & $f_{\text{ref}}$ & $e_{\text{ref}}$ & $l_{\text{ref}}$ & $\chi_{\text{eff}}$ & $\chi_p$ \\ \hline
1 	 & 	 EccContPrecDiff007 	 & 	 30 	 & 	 2.5 	 & 	 72 	 & 	 20 	 & 	 0.080 	 & 	 0.466 	 & 	 0.00 	 & 	 0.41 \\ \hline
2 	 & 	 ICTSEccParallel15 	 & 	 73 	 & 	 5.0 	 & 	 48 	 & 	 21 	 & 	 0.097 	 & 	 6.057 	 & 	 0.30 	 & 	 0.00 \\ \hline
3 	 & 	 ICTSEccParallel08 	 & 	 16 	 & 	 5.0 	 & 	 90 	 & 	 22 	 & 	 0.121 	 & 	 4.946 	 & 	 0.05 	 & 	 0.00 \\ \hline
4 	 & 	 EccContPrecDiff008 	 & 	 40 	 & 	 1.0 	 & 	 40 	 & 	 24 	 & 	 0.123 	 & 	 0.649 	 & 	 0.00 	 & 	 0.50 \\ \hline
5 	 & 	 ICTSEccParallel17 	 & 	 42 	 & 	 5.0 	 & 	 54 	 & 	 20 	 & 	 0.132 	 & 	 3.384 	 & 	 -0.55 	 & 	 0.00 \\ \hline
6 	 & 	 EccPrecDiff002 	 & 	 159 	 & 	 1.0 	 & 	 30 	 & 	 20 	 & 	 0.315 	 & 	 2.162 	 & 	 0.20 	 & 	 0.20 \\ \hline
\end{tabular}